\documentclass[prapplied,rsi, amsmath,amssymb,reprint,superscriptaddress]{revtex4-1}

\usepackage{longtable}
\usepackage{graphicx}
\usepackage{color}
\usepackage{soul}

\usepackage[version=3]{mhchem} % Formula subscripts using \ce{}
\usepackage{blindtext}
\usepackage{bm}
\usepackage{array}
\newcolumntype{C}{>{\centering\arraybackslash}p{4em}}
\newcolumntype{D}{>{\centering\arraybackslash}p{6em}}

\begin{document}
\title{Non-classical nucleation of zinc oxide from a physically-motivated machine-learning approach.}

\author{Jacek Goniakowski}
\affiliation{CNRS, Sorbonne Université, Institut des NanoSciences de Paris, UMR 7588, 4 Place Jussieu, F-75005 Paris, France}

\author{Sarath Menon}
\affiliation{Computational Materials Design, Max-Planck-Institut für Eisenforschung GmbH, 40237 Düsseldorf, Germany}

\author{Ga\'etan Laurens}
\affiliation{Institut Lumi\`ere Mati\`ere, UMR5306 Universit\'e Lyon 1-CNRS, Universit\'e de Lyon, 69622 Villeurbanne Cedex, France}

\author{Julien Lam}
\email{julien.lam@cnrs.fr}	
\affiliation{Centre d’élaboration des Matériaux et d’Etudes Structurales, CNRS (UPR 8011), 29 rue Jeanne Marvig, 31055 Toulouse Cedex 4, France}

\begin{abstract}
Observing non-classical nucleation pathways remains challenging in simulations of complex materials with technological interests. This is because it requires very accurate force fields that can capture the whole complexity of their underlying interatomic interactions and an advanced structural analysis able to discriminate between competing crystalline phases. HereWe first describe how we used the Physical LassoLars Interaction Potentials technique to create a machine-learning force field for zinc oxide interactions. Then, we carried out several types of crystallization simulations and followed the formation of ZnO crystal with atomistic precision. Our results, which were analyzed using a data-driven approach based on bond order parameters, demonstrate the presence of both prenucleation clusters and two-step nucleation scenarios, thus retrieving seminal predictions of non-classical nucleation pathways made on much simpler models. Dedicated calculations of high temperature ZnO free energy within a newly developed automated nonequilibrium thermodynamic integration method revealed the existence of a thermodynamic bias for the observed non-classical nucleation scenarios.
\end{abstract}

\maketitle

\section{Introduction}

While crystals in material science are ubiquitous, the mechanisms of their formation which spans from nucleation to crystal growth remain one of the most intriguing processes in nature. Better understanding crystallization would allow for a rational control of material engineering and possibly for the development of novel functional materials and technological applications. From the fundamental point of view, numerous works have been dedicated to elucidating the emergence of the nucleation core~\cite{TenWolde1995Oct,Wolde1997Sep,Auer2001Feb,Lutsko2020May,Lutsko2019Apr} and its role in controlling the final crystal structure~\cite{Sleutel2014Feb,VanDriessche2018Apr,Menon2020Sep,Lam2018Mar,Amodeo2020Oct,Desgranges2007Jun,Desgranges2006Nov}. For instance, it is now possible to observe the crystal birth with electron microscopy~\cite{Sleutel2014Feb,VanDriessche2018Apr,Yamazaki2017Feb,Sleutel2014Dec,Ogata2020Jan}, and colloidal science has also provided numerous experimental results on nucleation~\cite{vanBlaaderen1997Jan,Leunissen2005Sep,Velikov2002Apr,deNijs2015Jan}. Yet, numerical simulations remain the principal instrument to investigate crystallization at the atomistic level~\cite{Sosso2016Jun}. In this context, most of these works require large scale simulations in order to observe the phase transition and have therefore only focused on materials for which the interactions are very simple, including hard-spheres~\cite{Auer2001Feb,Coli2021Mar,Russo2012Jul,Leoni2021Jul,Pusey2009Dec,Sanz2011May}, Lennard-Jones~\cite{Trudu2006Sep,Lutsko2019Apr,TenWolde1995Oct,Lam2018Mar,Desgranges2007Jun}, water~\cite{Leoni2021Jul,Lin2018Nov,Metya2021Mar,Qiu2018Sep,Qiu2019May,Lupi2014Feb,Moore2011Nov,Lupi2017Nov,Russo2014Jul}, as well as metallic potentials, like embedded-atom model (EAM)~\cite{Menon2020Sep, Amodeo2020Oct,Desgranges2007Jun2}.

Prompted by this large body of fundamental achievements, it becomes timely to reach the same level of understanding for crystallization in more complex materials in order to target more diverse technological applications. So far, the need for large scale simulations have prevented from using quantum accurate modeling including density functional theory (DFT), and the research field dedicated to constructing novel interaction potentials to bridge this computational gap has been ever expanding~\cite{Combettes2020Sep,Laurens2020Jan,Calvo2017Mar,Zipoli2013Dec,Rajasekaran2016Mar,Senftle2016Mar,Brommer2015Sep,Vashishta2008Apr}. 
In particular, the past decade has seen the emergence of innovative types of interaction potentials based on machine-learning algorithms~\cite{Behler2007Apr}. Various approaches have been proposed such as Artificial Neural Networks~\cite{Behler2007Apr}, Gaussian approximation potentials~\cite{Bartok2010Apr}, Linearized potentials~\cite{Seko2015Aug,Seko2014Jul,Takahashi2017Nov,Seko2019Jun,Goryaeva2019Aug,Benoit2020Dec}, Spectral Neighbor Analysis Potential~\cite{Thompson2015Mar,Wood2018Jun}, Symmetric Gradient Domain Machine learning~\cite{Chmiela2017May,Chmiela2018Sep}, and Moment Tensor Potentials~\cite{Shapeev2016Sep,Novikov2020Dec}. The success of those machine-learning interaction potentials (MLIPs) is seen through the large variety of studied materials, namely pure metals~\cite{Novoselov2019Jun,Seko2015Aug,Takahashi2018Jun,Zeni2018Jun,Botu2017Jan}, organic molecules~\cite{Bereau2018Jun,Sauceda2019Mar,Bartok2017Dec,Veit2019Apr}, water~\cite{Nguyen2018Jun,Bartok2013Aug,Morawietz2012Feb,Natarajan2016Oct,Morawietz2013Aug}, amorphous materials~\cite{Deringer2017Mar,Bartok2018Dec,Caro2018Apr,Deringer2018Jun,Deringer2018Nov,Sosso2018Jul}, and hybrid perovskites~\cite{Jinnouchi2019Jun}.

The oxide materials has been much less investigated using MLIPs~\cite{Sivaraman2020Jul,Artrith2013Jun,Eckhoff2020Oct,Artrith2011Apr,Sundararaman2018May,Eckhoff2020Nov,Artrith2017Jul,Artrith2016Mar}.
With those materials, the complexity lies in the emergence of long-range electrostatic effects difficult to capture with traditional machine-learning approaches~\cite{Ko2021Jan,Ko2021Feb}. In addition, oxide materials often exhibit a rich structural landscape composed of numerous polymorphs~\cite{Navrotsky2008Mar,Machala2011Jul,Sponza2015Feb,Tsybulya2008Jan}. Being able to model those polymorphs with a unique set of atomic interaction remains challenging even for advanced machine-learning methods.

In this article we focus on modeling ZnO crystallization from bulk ZnO melt, based on a newly developed approach named Physical LassoLars Interaction Potential (PLIP), which employs a physically-motivated mathematical formulation and relies on a constrained linear regression scheme for parameter adjustment. Beyond the existing state-of-the-art interaction models, such as ReaxFF, Tersoff potentials, or artificial neural network potentials~\cite{Artrith2011Apr,Artrith2013Jun}, we explicitly show that our PLIP satisfactorily recovers DFT accurate results in numerous situations including the six most stable bulk ZnO polymorphs and their low index surfaces, the vibrational characteristics along with the  corresponding free energy behavior of the most stable ZnO crystalline phases, and the structural properties of amorphous ZnO and of its high temperature melt.

Having demonstrated the ability of the PLIP to model ZnO in all of these situations, we used it in molecular dynamics (MD) simulations of ZnO crystallization by either freezing ZnO melt or by letting it occur spontaneously in an undercooled ZnO liquid. With the help of a data-driven approach for phase recognition, we were able to access the nucleation of zinc oxide with an atomic precision and to reveal a dominant role of a non-classical nucleation scenarios, involving the less common BCT ZnO polymorph. Moreover, with dedicated calculations of ZnO free energy within a newly developed automated nonequilibrium thermodynamic integration method we showed the existence of a stability reversal between the most stable phase and the BCT polymorphs at high temperature thus proving that the predicted non classical nucleation pathways are mostly driven by thermodynamic processes.

From a material point of view, since finite size zinc oxide crystals are found in numerous applications, $e.g.$ photocatalysis, piezzoelectricity, drug delivery, antibacterial, and gas detection\cite{Primo2020Nov}, our results provide tools and lay grounds for future studies on more complex, finite size and surface- and/or interface phenomena which are involved in formation of ZnO nanostructures.

\section{Methods}

\subsection{Machine-learning interaction potential using the PLIP approach}

As in most MLIP approaches, the total potential energy $E$ is decomposed as the sum of independent atomic energies $E_i$: $E = \sum^{N}_{i=0} E_i$, where $N$ corresponds to the number of atoms of the considered configuration. Then, we employ a linear model which consists in approximating $E_i$ as a  linear combination of descriptors $X_n^i$: $E_i = \sum_{n} \omega_n X_n^i$ where $\omega_n$ are the linear coefficients that must be determined. More specifically, the PLIP model is made of three types of descriptors which explicitly follow a many-body order expansion:
\begin{align}
&[2B]_n^i = \sum_j f_n(R_{ij}) \times f_c(R_{ij}), \\
&[3B]_{n,l}^i = \sum_j \sum_k f_n(R_{ij})f_c(R_{ij}) f_n(R_{ik})f_c(R_{ik})cos^l(\theta_{ijk}),  \\ 
&[NB]_{n,m}^i = \left( \sum_j f_n(R_{ij}) \times f_c(R_{ij}) \times f_s(R_{ij}) \right)^m,
\end{align}
where $R_{ij}$ is the distance between atoms $i$ and $j$, $\theta_{ijk}$ is the angle centered around the atom $i$, and $l$ and $m$ are two positive integers. $f_c(R_{ij})$ is the same cut-off function as in the seminal work of Behler and Parinello~\cite{Behler2007Apr}, namely $f_c(R_{ij}) = \frac{1}{2} \left(1 + cos(\pi(R_{ij}/R_{cut}))\right)$, where the cut-off distance $R_{cut}$ is set at 6~\AA. The shift function is chosen in the from: $f_s(u) = 6u^5 -15u^4 + 10u^3$, where $u = (R_{ij} - r_1)/(r_2 - r_1)$, and $r_1$ (resp. $r_2$) is defined as 95~\% (resp. 105~\%) of a short distance equal to $1.1$\,\AA. Regarding the basis functions $f_n(R_{ij})$, we previously demonstrated the advantage of coupling different classical functions at the same time~\cite{Benoit2020Dec}. Yet, in this particular study on zinc oxide where numerical efficiency is crucial, we work only with Gaussian functions for which the width and the central positions are respectively listed as follows: [0.5, 1.0, 1.5] and [0.5, 1.0, 1.5 ... 5.5, 6.0]. In addition, we vary the integers $l$ and $m$ respectively from $0$ to $5$, and from $4$ to $7$. Altogether, when taking into account the binary nature of the zinc oxide system, our model is made of $1981$ available descriptors. 

In order to match the first-principle database, the model is fitted using the LassoLars approach which allows for a well-informed selection of the most preponderant descriptors and a reduction in the complexity of the obtained potential~\cite{Benoit2020Dec,Zeni2021Jun}.  In this study, each of the PLIP models selects approximately $150$ descriptors among an order of magnitude more that are available.

\subsection{First principle calculations and training database}
\label{DFT}

Reference GGA-DFT calculations are performed with VASP~\cite{KressePRB1993,KressePRB1996}, using PW91 exchange-correlation functional~\cite{PerdewPRL1996}, and the projector augmented wave method~\cite{KressePRB1999}. Standard zinc and soft oxygen (energy cutoff of 270 eV) pseudopotentials provided by VASP are used in all calculations, enabling an efficient structural relaxation of systems composed of several hundreds of atoms. GGA results obtained with the soft and the full (energy cutoff of 400 eV) oxygen pseudopotentials show satisfactory agreement (differences of bulk parameters $a$ and $c$ smaller than 0.01 \AA, and cohesion energy differences $\Delta E_\mathrm{coh}$ below 0.01 eV/ZnO). The present results on six ZnO polymorphs [wurtzite (WRZ), zinc blend (ZBL), body centered tetragonal (BCT), sodalite (SOD), $h$-BN (HBN), and cubane (CUB) crystallographic structures] coincide very well with the existing hybrid HSE03 reference~\cite{Sponza2015Feb} [See Tab. SI1], in particular concerning the relative stability of the considered polymorphs. Such agreement between the two types of modeling indicates that the chosen GGA-DFT approach leads to sufficiently accurate results while enabling for larger scale calculations. 

Since the training database includes both ordered and disordered structures, in DFT calculations we systematically use relatively large supercells (16-19 \AA~large cuboids), containing 320-480 atoms, and sample the Brillouin zone with a single $\Gamma$ point. In ordered structures, the atomic coordinates of all ions are relaxed until residual forces dropped below 0.01 eV/\AA~and, in the bulk structures, all components of the stress tensor are smaller than 0.01 eV/\AA$^{3}$. The disordered structures are collected along short high-temperature MD runs with no further relaxation as to produce non-vanishing forces on all ions. 

In order to include information on the characteristics of low coordinated ions, for each of the six ZnO polymorphs we also consider at least one low index non-polar surface, whereas three surface orientations are considered for the most stable (WRZ) polymorph. In this latter case, beyond the most stable (10-10) and (11-20) non-polar orientations, we also include the polar (0001)/(000-1) surfaces. As to impede the emergence of a macroscopic dipole moment, we use an asymmetric slab with the conventional (2$\times$2) reconstructed surfaces at which one surface oxygen (zinc) ion of each four is removed at the oxygen (zinc) termination~\cite{GoniakowskiRPP2007}. As a consequence, in this case the calculated surface energy corresponds to an average value of the zinc- and oxygen-terminated surfaces. In all calculations, slabs of 6-12 atomic layers are separated by about 15 \AA\ of vacuum, and the atomic coordinates of all ions are fully relaxed. The calculated surface energies, Tab. SI2, are fully consistent with the existing computational evidence~\cite{ClaeyssensJMC2005,FreemanPRL2006,GoniakowskiPRL2007,MorganPRB2009,DemirogluPRL2013,Sponza2015Feb}.

Three PLIPs are developed with different training datasets. Indeed, we train a first PLIP on a database composed of the 6 bulk polymorphs melted up to 5000\,K. For those first MD runs, we use a classical Buckingham ZnO potential\cite{Binks1993Sep} which was previously employed in numerous studies of ZnO\cite{Wang2010Jan,Lin2013Feb,Kulkarni2005Oct,Kulkarni2006Sep}. We would like to emphasize that using a classical potential instead of first-principle calculations at this stage favors a more rapid sampling of a larger variety of configurations. Then, for the second database, we considered ZnO surfaces heated up to 2000 K using the first PLIP model as well as structures of the first database. Finally, for the third model, additional  amorphous bulk obtained after rapidly cooling liquid structures from 2000\,K to room temperature are added to the previous database. There, we employed the second PLIP model for the interaction potential. During the MD runs, we extracted respectively 20 and 10 for the bulk and the surface structures as well as 15 more amorphous structures. Thus, we sequentially increase the size and the variety of the database always using the previous model for sampling. Those three PLIPs are respectively denoted V1, V2, and V3 in the following. To validate the PLIPs, we perform a series of calculations where bulk and surface properties are computed for the six ZnO polymorphs, and compared to DFT results. We employ the conjugate gradient algorithm, with a force convergence criterion of 10$^{-9}$ eV/\AA. Moreover, we also investigate the efficiency of the PLIPs in the context of disordered ZnO structures, such as liquid and amorphous phases. For this purpose, simulations are carried out by starting with 500 atoms randomly disposed and heated at 3000 K, and then cooled to 300 K with a rate of 1800 K/ns. 

\subsection{Molecular dynamics simulations of crystallization}

After developing the PLIP models, we perform two complementary types of molecular dynamics simulations of crystallization.  

On the one hand, in freezing simulations the temperature is progressively lowered as to initiate the crystallization. In this case, the system is initialized with 1000 randomly positioned atoms of zinc and the same number of oxygen atoms. Then, atomic positions are first optimized to remove large forces due to overlapping atoms. We further proceed to a first equilibration at 5000 K during 1 ps in the NVT ensemble with a Nose-Hoover thermostat. From preliminary freezing simulations (not shown) we  found out that within the considered conditions crystallization occurs in the temperature range of 1500 K to 1000 K. Therefore, in the reported simulations, the freezing is operated in two steps. In the first step, the system is quenched from 2500 K to 1500 K during 100 ps which allows us to quickly obtain a first liquid structure. In the second step, the system is slowly cooled down from 1500 K to 1000 K during 10 ns. These two cooling steps are both performed in the NPT ensemble at P = 1 bar. 

On the other hand, we also performed constant temperature simulations. For that purpose, liquid structures are generated just as in the freezing simulations. Then, a thermostat and a barostat are employed to maintain the temperature and pressure until spontaneous crystallization is observed. 

The Large-scale Atomic/Molecular Massively Parallel Simulator (LAMMPS) package~\cite{Plimpton1995Mar} with a custom-made pair style for PLIP modeling is used in all the simulations.

\subsection{Data-driven approach for the structural analysis}

Analysis of the obtained structures and the crystallization dynamics requires the use of novel order parameters which are able to distinguish between each of the six crystal polymorphs of zinc oxide. Such a challenging task is clearly out of the reach of classical order parameter including Common neighbor analysis, Voronoi cell topology, or Polyhedral template matching which are mostly designed for monodisperse systems and for classical crystal structures like face-centered cubic and body centered cubic. In the case of zinc oxide crystallization, the very rich structural landscape requires a structurally agnostic order parameter, namely one that is transferable for all kinds of crystal structures. 

\begin{figure}[!ht]
  \includegraphics[width=8.6cm]{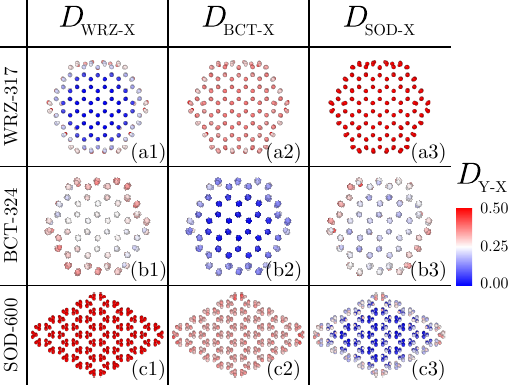}
  \caption{Structural analysis of three different nanoparticles obtained from Viñes et al.~\cite{Vines2017Jul} that are respectively in the WRZ (a), BCT (b) and SOD (c) crystal phases. We measured the distances with respect to the same three different crystal phases in bulk.}
  \label{TestQ}
\end{figure}

In this context, we develop a data-driven approach for the structural analysis. In practice, averaged Steinhardt parameters denoted $\overline{q_l}$ with $l\in[3:6]$ are computed for zinc and oxygen atoms when considering the entire system (zinc and oxygen atoms) and only inter-species distances (Zn-Zn and O-O). We use the \texttt{pyscal} Python package~\cite{Menon2019Nov} with the adaptive methods to compute the neighbor lists. This calculation leads to a Steinhardt vector denoted $\bm{Q}$ of $12$ components (6 for the entire system and 6 for inter-species distances) for each atom. Then, we compute the same vector in the six bulk polymorphs to construct a reference database averaged over all the zinc and oxygen atoms in those crystal structures. If a studied atom has the same local structure as one in the bulk polymorph, its Steinhardt vector must be close to that of the database which can be measured using Euclidean distance in vector space denoted $D_{Y-X}$. In Fig.\,\ref{TestQ}, we test those Euclidean distances on nanoparticles of known structures and show that the method is able to identify the correct structure. In addition, the method retrieves that the center of the nanoparticle is more similar to the bulk than its edges. 

Finally, in order to quantify these Euclidean distances, we compute a likelihood parameter denoted $S$ measured along a path between two reference structures named A and B using:
\begin{equation}
S=\frac{e^{-\lambda D_{AX}}+2e^{-\lambda D_{BX}}}{e^{-\lambda D_{AX}}+e^{-\lambda D_{BX}}}
\end{equation}
where $\lambda$ is adjusted so that $D_{AB}=2.3/\lambda$. As such, we obtain a score that quantifies the local structure: an atom which is similar to structure A (resp. B) gives $S\sim 1$ (resp. $S\sim 2$). Therefore, we consider that the value of $1.5$ discriminates between the two types of structure. 

In retrospective, our approach combines the use of (1) a vector of Steinhardt parameters which was already suggested for colloidal systems and amorphous models~\citep{Boattini2020Oct,Boattini2019Oct}, (2) a similarity measurement with the identity function as Kernel, and (3) a path collective variable which was previously employed for the construction of order parameters based on permutation invariant vectors~\cite{Pipolo2017Dec}.

\section{Results}

\subsection{Validation of the PLIP models}

\subsubsection{Bulk crystals}
\begin{figure}[h!]
\includegraphics[width=8.6cm]{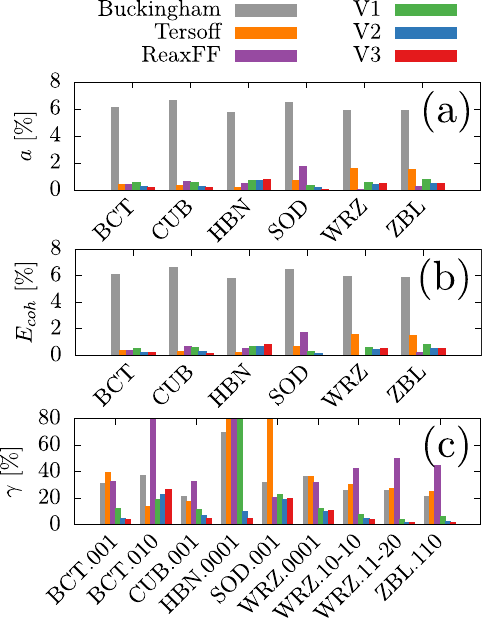}
\caption{Absolute errors (\%) with respect to the DFT values of bulk lattice parameter $a$, cohesion energy $E_{coh}$ relative to the wurtzite polymorph, and the relative surface energy $\gamma$ determined from the relaxed structures obtained with the three PLIPs as well as with three classical potentials. For signed values of the error, please refer to Fig. SI1.}
\label{fig:crystal}
\end{figure}

As a first step to validate the developed PLIPs, the crystal structures of the six polymorphs were optimized using the three PLIPs, V1, V2, and V3. The resulting lattice parameter $a$ and cohesion energy $E_{coh}$ relative to the lowest-energy WRZ phase were compared to those obtained using DFT in order to evaluate the accuracy of the PLIPs. Errors with respect to DFT values for these two quantities are displayed in Fig.~\ref{fig:crystal}.(a) and (b), respectively, along with results obtained with three classical force-field potentials, namely the Buckingham~\cite{Binks1993Sep}, Tersoff~\cite{Erhart2008}, and ReaxFF~\cite{Raymand2008} potentials. 

Regarding the three PLIPs, they give approximately the same results, with the V2 and V3 PLIPs being slightly more accurate than V1. The already accurate results obtained with V1 indicate that training with only bulk ZnO structures is sufficient to reach good model for the bulk crystals. Except for the Buckingham potential for which the errors are considerably higher, the two other classical potentials present comparable errors for all the polymorphs with respect to the PLIP models.

Accuracy on ZnO surfaces was also investigated and the errors on the surface energy are computed both using PLIPs and the classical potentials [See Fig.~\ref{fig:crystal}.(c)]. When comparing each PLIP version, we note that while V2 and V3 are accurate by construction, V1 in most cases leads to very good results in terms of surface energy although it was not trained with any surface structures. Contrary to bulk parameters, the classical potentials are less accurate at predicting the ZnO surface energies compared to the PLIPs.  Surprisingly, the Buckingham potential, while the simplest in its mathematical formulation, exhibits the best overall results among the classical potentials. However, the PLIP errors are generally much lower than what is obtained with any of the classical potentials with an improvement of a factor between 3 and 8.5.

\begin{figure}[h!]
\includegraphics[width=8.6cm]{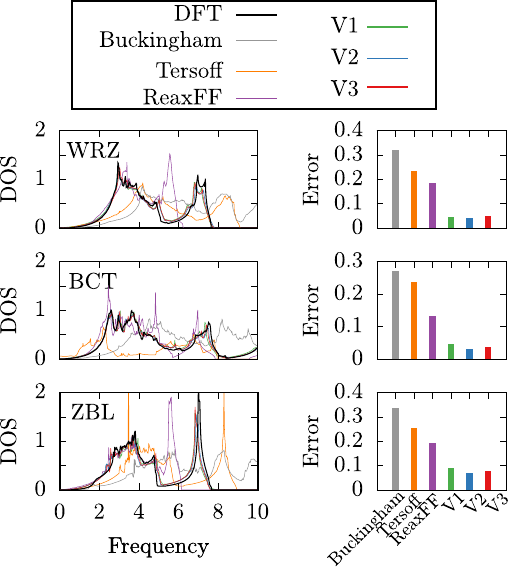}
\caption{Phonon density of states in three crystal structures. Quantitative comparison is calculated as the difference of the obtained curves with respect to the DFT reference.}
\label{fig:DOS}
\end{figure}

In order to further test the accuracy of PLIPs on bulk crystals, we computed the phonon density of states (DOS) for the three most stable crystal structures ie. WRZ, ZBL and BCT. For that purpose, the phonon frequencies are computed by diagonalizing of the dynamical matrix obtained through measuring the variation of atomic forces due to finite atomic displacements. To increase the calculation accuracy, we expanded the conventional unit cells following: WRZ (5$\times$5$\times$3), ZBL(3$\times$3$\times$3) and BCT (3$\times$3$\times$5). The phonon calculations were performed using the PHONOPY code\cite{Togo2015Nov,Togo2008Oct}. From Fig .\,\ref{fig:DOS}, one can see that although the tested classical potentials provided acceptable results for static properties like lattice parameters and cohesive energies, they are extremely inaccurate for dynamical properties as shown by the DOS. From a qualitative picture, ReaxFF which was the most accurate classical potentials already shows additional peaks in the DOS especially in the case of WRZ and ZBL. Quantitatively, the error of the three PLIPs is at least two times smaller than any of the classical potentials. Along with the previous measurement on surface energy, this results using DOS provides an additional argument for the necessity of constructing PLIP models for the zinc oxide system.

Finally, we note that we evaluated the melting temperature using the phase coexistence approach and obtained $1750 \pm 50$\,K for the PLIP V3 model which is about $500$\,K lower than the experimental value. The excellent matching of PLIP and DFT vibrational characteristics for the key ZnO crystalline phases [See Fig.\,\ref{fig:DOS}], suggests that this discrepancy originates from deficiencies not of the PLIP itself but of the DFT approximation.

\subsubsection{Disordered regimes}

\begin{figure}[h!]
\includegraphics[width=8.6cm]{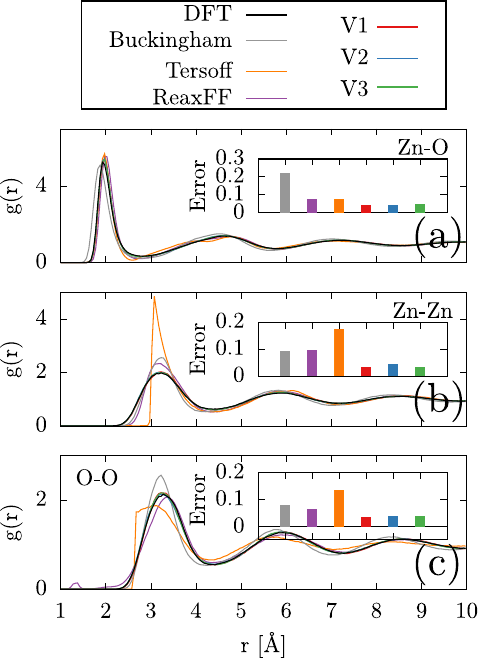}
\caption{Partial radial distribution functions $g(r)$ of the liquid structures obtained at 1500\,K. Structures calculated with the three PLIPs are compared with those computed using DFT and the three potentials. Quantitative comparison is obtained with the absolute errors displayed in insets and calculated as the difference of the RDF curves with respect to the DFT reference.}
\label{fig:liquid}
\end{figure}

After the crystal properties of ZnO, we were interested in measuring the accuracy of the model in the case of disordered  structures. For this purpose, we performed ab initio MD simulations using three different initial amorphous structures in the NVT ensemble at 1500\,K. The total duration of each simulation was equal to $4$\,ps with a timestep of $1$\,fs. The obtained structures, were analyzed by calculating the partial RDFs. In Fig.~\ref{fig:liquid}, we display results obtained with the DFT calculations, classical potentials and PLIP models. Quantitative errors based on the difference between RDF curves and DFT ones are also indicated for a better comparative evaluation. The RDFs coming from the PLIP structures fit perfectly those of DFT structures (visually and with the error measurements), while a mismatch is clearly visible between the DFT and the classical potential curves. Indeed, the main peak of the Buckingham potential curves is misaligned in all the partial RDFs, and both Tersoff and ReaxFF curves are completely different for the two homonuclear curves. In the case of O-O, the ReaxFF and Tersoff curves also predict an erroneous small peak at around 1.3~\AA. Regarding the errors, the PLIPs are undoubtedly closer to the DFT results than al of the classical potentials, with errors generally 2 to 3 times smaller. Moreover, even if all the PLIP RDFs are similar, a slight advantage is given to the V3 curve, especially at the first peak. It remains that the training of V1 and V2 models on hot liquid structures seems sufficient to describe the amorphous structures stabilized at room temperature. Results on the bond angle distributions shown in Fig. SI3 also displays a much better agreement with DFT calculations for all the PLIP models. We would like to emphasize that even if none of the DFT liquids were included in the database, the PLIP models are still able to extrapolate towards such high temperatures regime. 

Prompted by those results on liquid structures, we further tested the obtained PLIP in the amorphous regime. In particular, To generate amorphous structures, a hot liquid was cooled until reaching the ambient temperature with a cooling rate equal to $1.8 \times 10^6$\,K/ns which allows for the formation of the amorphous form of ZnO. Finally, the obtained structures were further optimized in energy and forces down to respectively $10^{-4}$\,eV and $10^{-6}$\,eV/\AA. As such, twenty different amorphous structures made of 360 atoms were constructed using different initial seeds and were analyzed by calculating the partial RDFs. In Fig.SI4 and SI5, we display results obtained with the DFT calculations, classical potentials and PLIP models. Results are very similar to what we obtained in the liquid regime. Indeed, while more simple in its mathematical formulation, the Buckingham potential is generally the best among the classical force field with both Tersoff and ReaxFF showing nonphysical behavior respectively for the Zn-Zn and the O-O first neighbor. Then, all of the obtained PLIP exhibit a very good agreement with DFT calculations.

\subsubsection{Nanostructures}

\begin{figure*}
  \includegraphics[width=\textwidth]{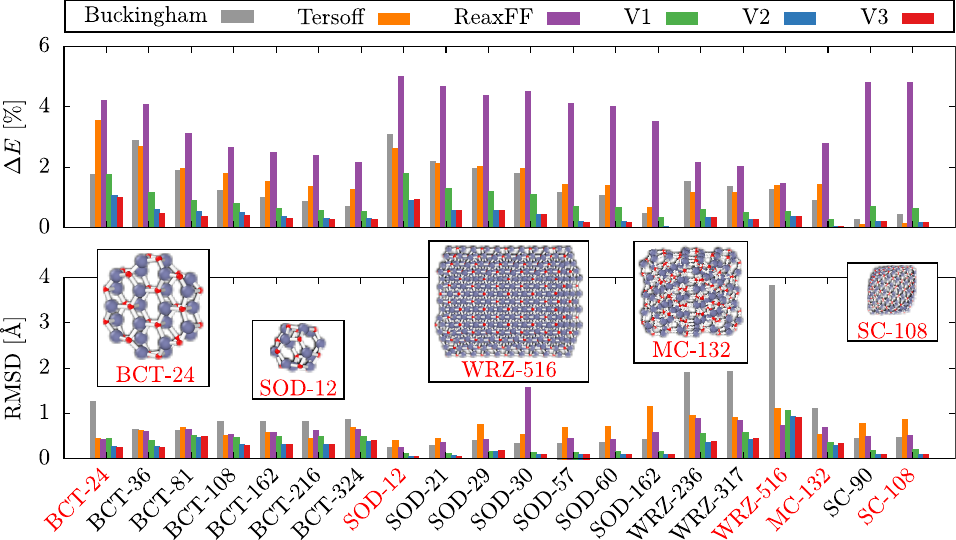}
  \caption{(Top panel) Relative absolute error on the clusters energy (per \ce{ZnO} formula unit) relative to the Wurtzite bulk energy obtained with the PLIPs and the classical potentials. (Bottom panel) Corresponding RMSD based on comparison of atomic positions with those in DFT configurations obtained from structural optimization of geometries published by Viñes et al.~\cite{Vines2017Jul}. Typical clusters are represented in the insets. For signed values of the errors on the energies, please refer to Fig. SI2.}
  \label{Nano}
\end{figure*}

For this last test of the PLIP models, we employed clusters of ZnO obtained by Viñes et al.~\cite{Vines2017Jul} using DFT. They focused on three polymorphs, namely BCT, WRZ and SOD, and also worked with single-caged (SC) and multi-caged (MC) structures and reported the most probable low energy structures for systems made of tens to hundreds of atoms. In our test, we additionally optimized these structures using our own DFT flavor, the PLIP models and the three potentials. With the obtained structures, we computed the formation energy difference $\Delta E$ with respect to DFT results. Moreover, as to asses the quality of cluster structures geometries, we measured the root-mean-square deviation (RMSD) based on the difference of atomic positions with respect to the DFT reference and extracted the maximum value obtained when examining every atom of the nanostructure. Fig.\,\ref{Nano} shows that the classical potentials are not able to provide accurate results on these nanostructures while the three versions of PLIP give reliable results when compared to DFT calculations. 

Interestingly, the PLIP models are able to accurately reproduce the energy and the geometry even for the smallest clusters that are qualitatively different from the bulk-like structures of the largest clusters, which is an evidence for their transferability towards untrained structures. In particular, the single cage SC-$108$ and SC-$90$ do not show any bulk-like structure since these are shell clusters, and the PLIP models still give very small errors. For the three PLIPs, a slight advantage is still given for the V2 and V3 models for which the $\Delta E$ and RMSDs do not exceed 1~\% and 0.5~\AA, respectively. 

\vspace{0.5cm}

Altogether, it is worth noticing that although the PLIP models were only trained with DFT forces, they are still able to reach very accurate results for energetic properties (bulk, surface, and nanoparticles). From now on, results presented in the remaining article will consider only the calculations having employed the V3 potential because it is built with a more complete database.

\subsubsection{Crystallization differences}

\begin{figure}[h!]
\includegraphics[width=8.6cm]{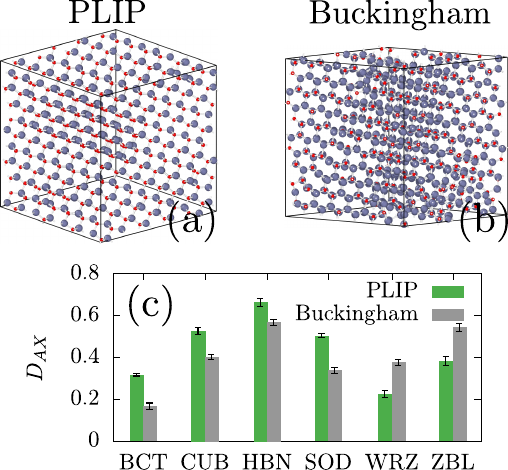}
\caption{Structures obtained after freezing simulations using (a) the PLIP model and (b) the Buckingham potential, after a slow cooling during $10$\,ns. (c) Euclidean distances between the obtained structures and the bulk perfect crystals computed in the Steinhardt vector space. Results are averaged over 5 simulations obtained with different initial conditions.}
\label{END_BuckVsMLIP}
\end{figure}

In this last validation section, we focus on crystallization results. In particular, we performed the exact same freezing simulations using PLIP as well as the 3 classical force fields. First, we note that the Tersoff and the ReaxFF models do not lead to crystallization under the considered thermodynamics conditions. Indeed, in both cases, the difficulty in modeling the oxygen-oxygen short range interactions already raised in case of the disordered ZnO leads to a nonphysical behavior in the high temperature regime where freezing simulations are started. It is possible that those two classical potentials may also lead to crystallization under more finely tuned initial conditions, including, e.g., the initial distribution of atoms and temperature of the freezing runs. However, one may also expect that the erroneous description of the short range oxygen-oxygen interaction impacts  the nucleation and the initial ZnO growth, where small grains are in contact with the liquid. Then, Fig.\,\ref{END_BuckVsMLIP}.(a,b) shows the structures obtained at the end of the freezing simulations with PLIP and the Buckhingham model. A quantitative picture is obtained by using the Euclidean distances in Steinhardt vector space between the final structures and those in the bulk database. From Fig\,\ref{END_BuckVsMLIP}.c, it appears that the smallest distance for the PLIP results is obtained with the WRZ structure while for the Buckingham potential, it corresponds to the BCT structure. This quantitative observation suggests that this classical potential leads to a qualitatively different result, $i.e.$ the growth of a BCT structure instead of the Wurtzite one, which is the most stable in both experiments and DFT calculations. This surprising result may be in part assigned to the fact that the classical potential overestimates stability of the BCT phase. Indeed, according to bulk DFT results the BCT phase is less stable than the WRZ one by $0.048eV/ZnO$ but this energy difference is reduced to $0.017eV/ZnO$ only within the Buckingham potential. In contrast, the PLIP model is able to retrieve a much more satisfying value of $0.044eV/ZnO$. This analysis on the final structure obtained after freezing shows that PLIP is able to crystallize the most stable structure which is a strong justification for using the PLIP model over a classical potential.

Thus, one of the key advantages for using the PLIP model over classical interaction is that, along with the quantitatively better results for bulk, surface, disordered and nanoparticle properties, it also leads to qualitatively better outcomes of freezing simulations. In addition, when comparing to DFT, the main advantage is the possibility for larger scale simulations. This is crucial in studies of crystallization where freezing simulations have to cover more than a few nanoseconds because shorter ones result in amorphous structures. Furthermore, much larger unit cells allowed by our PLIP help avoiding any nonphysical behavior due to interactions between nucleation cores in neighboring cells.

\subsubsection{Computational efficiency}

\begin{figure}[b!]
  \includegraphics[width=8.6cm]{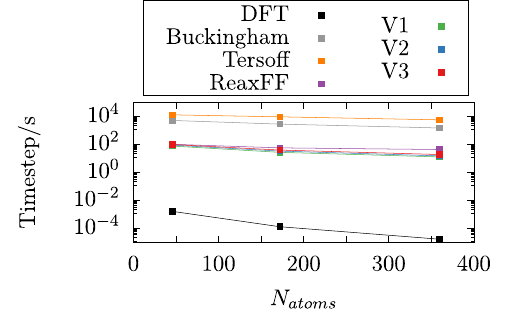}
  \caption{Computational efficiency of different models of ZnO interactions measured as the number of timesteps that can be reached in a second. Results were obtained on 16 CPU cores of Intel Skylake 6140 2.3\,GHz. } 
  \label{CompTime}
\end{figure}

Before closing on this validation section, Fig.\,\ref{CompTime} shows the computational efficiency of different models of ZnO interactions. It appears that the PLIP potentials are between 4 and 6 orders of magnitude faster than DFT depending on the system size. While being visibly slower than the most simple potentials (Buckingham and Tersoff) their efficiency is comparable to the that of the comparatively less reliable ReaxFF model. Furthermore, we would like to quickly comment on how easy the methodology can be applied to other systems. A first bottleneck is the construction of the database. In our case, a sequential increase of the database size was carried out by using first a classical empirical potential and then the previous version of the machine-learning force fields. As such, this part of the methodology is not as CPU consuming as the usage of ab initio MD for the database sampling and it still allows us to rapidly obtain structures that are highly uncorrelated. In addition, a second key aspect of the overall methodology is the computational cost of the learning process. In our case, because we employ LassoLars which is a constrained linear regression methods, we can obtain a PLIP model in approximately 10 minutes using 16 CPU cores which is relatively a low expense compared to more advanced non-linear regression methods.

\subsection{Temporal evolution during the ZnO crystallization}

\begin{figure*}[t!]
\includegraphics[width=\textwidth]{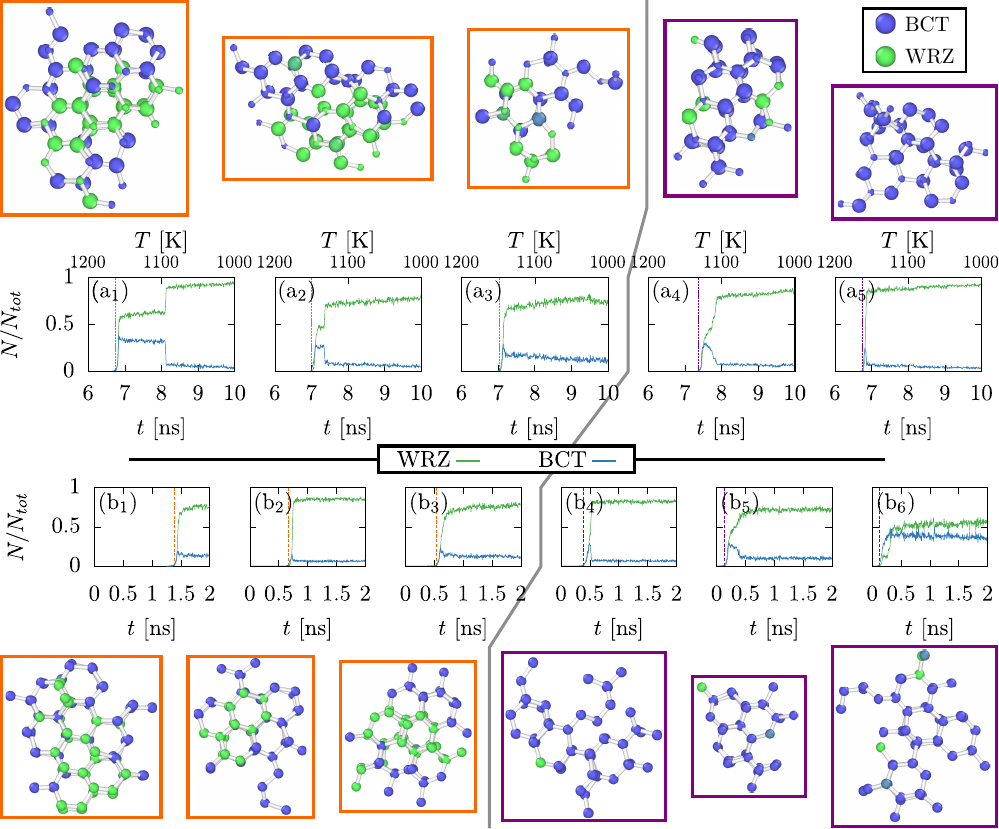}
\caption{Number of atoms detected in WRZ or BCT phases, according to the likelihood parameter $S$, along the crystallization path for independent MD runs using 2000 atoms and initiated with different atom positions and velocities. Top panels: freezing simulations at cooling rate equal to 50\,K/ns. Bottom panels: constant-temperature simulations at T = 1150 K. The red solid line discriminates between the two observed nucleation processes. Images of the typical nucleating structures show two nucleation scenarios: (A) WRZ surrounded by BCT (orange) and (B) Mostly BCT (violet).}
\label{VsTime}
\end{figure*}

Following the section dedicated to the validation of the PLIP model, we will now describe results obtained for ZnO crystallization, and the insights they give into the nucleation scenarios for this material. Two complementary types of crystallization simulations were performed with settings described in details in the Methods Section.

\subsubsection{Global observations}
In freezing simulations the ZnO liquid was slowly cooled down from 1500 K to 1000 K during 10 ns as to initiate crystallization. Results of five individual runs performed under the same thermodynamic conditions but with different initial atom positions and velocities are shown in Fig. \ref{VsTime} (top). To complement these out-of-equilibrium simulations, we have also mimicked infinitely slow cooling rates with constant temperature MD runs. Figure \ref{VsTime} (bottom) reports the results obtained in six runs at T = 1150 K, initiated with different atom positions and velocities. 

Regarding the structural evolution along the crystallization path, the behavior of the numbers of atoms in the predominant WRZ (green curve) and BCT (blue curve) environments reveals three growth stages in all the preformed runs: (1) an abrupt and quick increase of $N_{WRZ}$ and $N_{BCT}$ marks the nucleation and growth of WRZ and BCT  structures from liquid ZnO; (2) a plateau more or less broad indicates the stabilization of coexisting BCT and WRZ phases; (3) a phase transformation of BCT into WRZ visible by a drop of the curve of the former, compensated by the increase of the latter. We stress that similar results are obtained in freezing simulations with even a faster cooling rate of 25\,K/ns (See Fig. SI4), with a larger periodic cells (4000 atoms) (See Fig. SI5) and at different constant temperature simulations (T = 1100 K and 1200 K) (See Fig. SI6). This confirms that the observed nucleation and early ZnO growth characteristics are not biased by applied cooling rates (which are faster than these in most experimental situations) and are also not entangled by finite size effects.

Furthermore, Fig.\ref{VsTime} shows typical pictures of the nucleation core during both the freezing and the equilibrium simulations. As such, we find two types of structures for the nucleating clusters: (A) WRZ atoms surrounded by BCT phase and (B) Mostly BCT atoms. Among the reported results of 11 runs, we obtain 6 and 5 cases of scenario A and B, respectively. In the following, we will particularly focus on what happens during the first growth stage and we will further describe the two types of scenarios, exemplified in Figs. \ref{CaseAB}.

\subsubsection{Focus on the nucleation mechanisms}

\begin{figure*}[t!]
\includegraphics[width=\textwidth]{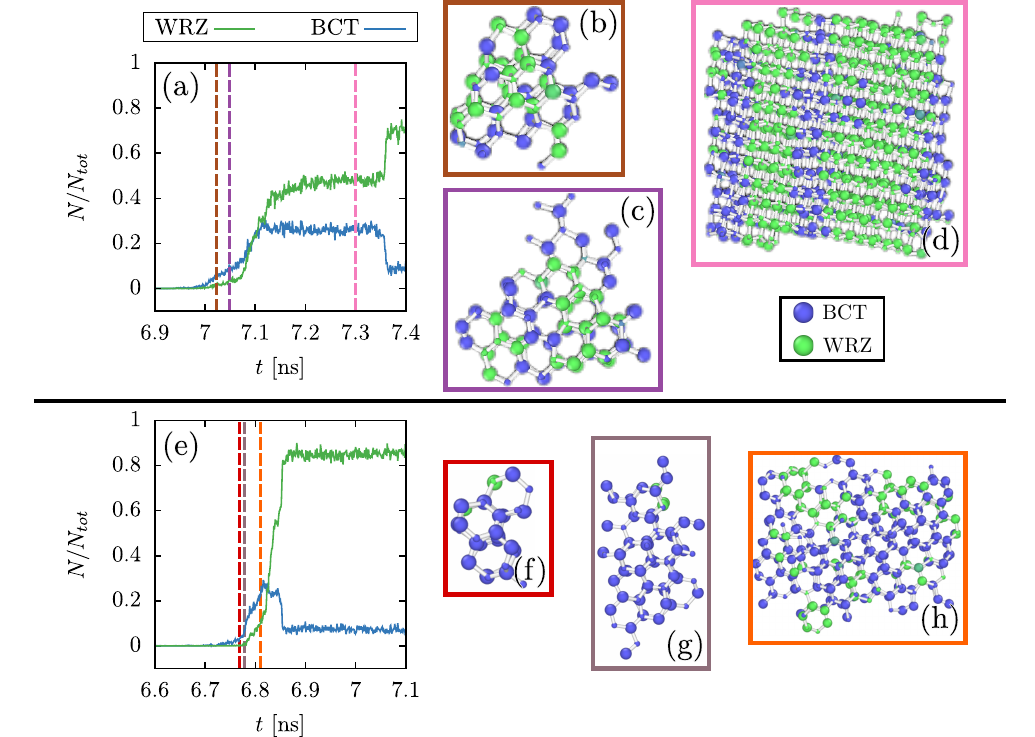}
\caption{Number of atoms detected in WRZ or BCT environments along the cooling path in first (top) and second (bottom) scenario which correspond respectively to Fig.\ref{VsTime}(a$_1$,a$_5$). Colored dashed lines point out where the structures are picked along the simulations. Atoms identified as WRZ and BCT according to the likelihood parameter $S$ are colored in green and blue, respectively.}
\label{CaseAB}
\end{figure*}

In the first scenario, Fig. \ref{CaseAB} (top), the growth is initiated by a nucleation of a WRZ grain. The behavior of $N_{WRZ}$ at the very beginning of the MD simulation, Fig. \ref{CaseAB}.a, shows that several small WRZ ZnO clusters form and dissolve in the ZnO liquid before the actual growth starts. From a structural point of view, the stability of the first grain seems to be enhanced by the formation of three-membered ZnO rings [See Fig.\,\ref{CaseAB}.b], characteristic of the Wurtzite (0001) surface. Indeed, in absence of such rings, similarly-sized ZnO clusters tend to dissolve. Most interestingly, the emerging WRZ clusters are systematically surrounded by a BCT phase taking the form of relatively thin and less crystallized structures [See Fig.\,\ref{CaseAB}.c]. Then, during the growth, the grain becomes significantly elongated, with virtually parallel (0001) facets at its opposite sides thus suggesting that crystallization along the (0001) atomic planes is favored at this growth stage [See Fig.\,\ref{CaseAB}.d]. Altogether, the first nucleation scenario comprises: (1) Stochastic appearance of unstable WRZ clusters that do not exhibit three-membered rings, (2) Emergence of a stable WRZ cluster made of three-membered rings and surrounded by a shell of BCT phase, and (3) Continuous growth of WRZ with a reminiscent BCT shell. 

In contrast, the second scenario, Fig.\,\ref{CaseAB} (bottom), shows quite a different nucleation path. Indeed, while $N_{WRZ}$ and $N_{BCT}$ increase quasi-simultaneously in the first scenario, here, the BCT phase emerges sooner than the WRZ one. Figs.\,\ref{CaseAB}.(f,g) clearly show that the emerging crystal is initially made of mostly BCT phase. Only once the BCT crystal is large enough, a solid-solid transition is observed  with BCT turning locally into WRZ. As a consequence, the BCT structure becomes porous, with multiple embedded WRZ clusters and is not stable enough to continue persisting. It is remarkable that while the initial BCT structure is much larger than in the first scenario, it can not survive as long as in the previous scenario because of this growth of embedded WRZ grains. The second scenario of nucleation can be summarized as a two-step mechanisms: (1) Metastable BCT crystals constitute the nucleation core and (2) The stable WRZ crystal only appears within large BCT crystals. An interesting feature of this second scenario is that we no longer observe growth of elongated WRZ grains since WRZ phase originates from multiple embedded smaller clusters. The elongated grains should therefore be favored by the presence of BCT shells.

\section{Discussions}

Both these scenarios highlight the essential role played by the metastable BCT phase in ZnO nucleation and early growth stages. Indeed, either it appears in the ZnO liquid simultaneously with the stable WRZ phase in form of a shell surrounding the growing WRZ clusters, or it nucleates first and the embedded WRZ clusters appear inside the metastable BCT matrix as a result of a solid-solid transition. From a fundamental perspective, both of these mechanisms contradict the classical nucleation scenario in which the first emerging crystal is made entirely of the most stable phase. Such non-classical nucleation scenarios have been already observed in simulations for model systems described by simple hard spheres\cite{Pusey2009Dec,Sanz2011May} and Lennard-Jones interactions\cite{Trudu2006Sep,Desgranges2007Jun} but, to our knowledge, the present prediction is among the first ones for systems with more complex, iono-covalent bonding thus confirming seminal experimental observations.\cite{Jehannin2019Jul,Lee2016Jun,DeYoreo2020}

\begin{figure}[b!]
  \includegraphics[width=8.6cm]{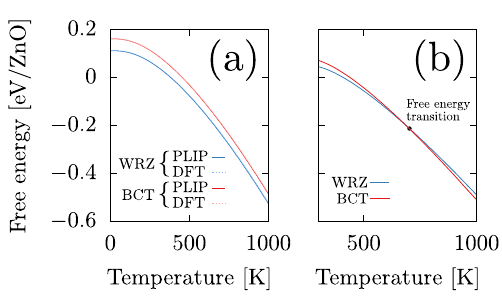}
  \caption{Free energy curves obtained under the harmonic approximation in DFT and PLIP V3 (a) and using the automated nonequilibrium thermodynamic integration method and the PLIP V3 (b).} 
  \label{FreeEnergy}
\end{figure}

While a thorough analysis of physical reasons of BCT involvement in the nucleation scenarios goes beyond the scope of the present study, it is worth pointing out that thermodynamic and kinetic effects may tend to interplay. On the one hand, from a kinetic viewpoint, the involvement of BCT phase in ZnO crystallization may results from a kinetic preference for an attachment pattern of \ce{ZnO4} tetrahedral units during crystal growth which favors the BCT phase. On the other hand, from a thermodynamic viewpoint, taking into account the small stability difference at 0\,K between the WRZ and BCT phases (0.044 eV/ZnO), the effect of lattice vibrations of differently arranged tetrahedral \ce{ZnO4} units in the two lattices (corner-connected only in WRZ, whereas some units are also edge-connected in BCT) may impact the relative stability of the two polymorphs at higher temperatures. In fact, our constant-temperature simulations, Fig. SI6, tend to show that the tendency to form long-lasting BCT increases with the temperature, which may suggest that BCT is less unstable at higher temperature.  In order to better understand the thermodynamic influence over the crystallization results, we first used the phonon dispersion curves shown in Fig.\,\ref{fig:DOS}.a to deduce the free energy in the harmonic approximation. Fig.\,\ref{FreeEnergy}.a shows that PLIP model is able to retrieve DFT values for the free energy curves which advocate furthermore for the accuracy of our MLIP approach. However, we must note that these calculations were obtained under the harmonic approximation and as such they can not be used to investigate the thermodynamic picture at high temperatures. To go beyond the harmonic approximation, we use a newly developed automated nonequilibrium thermodynamic integration method\cite{Menon2021Oct}.  In particular, we constructed two BCT and WRZ crystals made of 2880 atoms. We used thermodynamic integration going from the Einstein crystal to the PLIP model during 50 ps. Three cycles are carried out to determine the error which is equal to $0.004$\,eV/ZnO. Because of its large computational cost, such calculations are beyond the reach of DFT calculations but the PLIP excellent agreement with DFT within the harmonic approximation supports the extrapolation towards this other regime. Fig.\,\ref{FreeEnergy}.b shows that at such higher level of approximation, it exists a stability reversal between WRZ and BCT phases as the temperatures increases. This points towards thermodynamics as one of key factors at the origin of the reported non-classical nucleation mechanisms.

Further understanding of this thermodynamic picture would involve the measurement of the solid–liquid interfacial energies for both BCT and WRZ as well as rare-event sampling to compute the free energy barriers between the liquid phase and both WRZ and BCT crystals and also between the two crystals. However, the appropriate choice of collective variables for free energy calculations is already a fundamental challenge because of the very complex structural landscape. Altogether, these calculations require an enormous computational effort that can hardly be reached even with the PLIP model. 

Unfortunately, regarding experimental findings, the existing evidence for the bulk BCT phase of ZnO is relatively small since most studies focused on its appearance in ZnO ultra-thin films or axially strained nano-objects. Indeed, the BCT phase has been systematically predicted as the most stable ZnO polymorph in the thin film geometry\cite{Morgan2009Nov,Demiroglu2013Jun}, due to the more favorable energies of its low index surfaces which, in contrast to the basal WRZ(0001) surface, are non-polar. Also, a reversible WRZ-BCT reconstruction in the outmost layers of ZnO has been observed at some of ZnO surfaces.\cite{He2012Jul} These findings suggest that the shape/morphology of the crystallizing ZnO grains may also be an important factor.

An analogy can still be drawn with the case of titania where rutile and anatase phases tend to compete as observed with BCT and WRZ in zinc oxide.\cite{Manuputty2019} Anatase is considered as a metastable phase which can be kinetically stabilized at lower temperatures, owing to its less constrained structure and consequently enhanced kinetics of formation\cite{Hanaor2011Feb}. It has been pointed out that the more rapid crystallization of anatase may also be due to the lower surface free energy of this polymorph compared to that of rutile\cite{Ranade2002Apr}. From a computational viewpoint, zero-temperature DFT-based calculations predicted bulk anatase to be more stable than rutile and it is only most recently, that fine Diffusion Monte Carlo (DMC) simulations has suggested that bulk rutile may become more stable than anatase at higher temperatures, where the effect of lattice vibrations becomes crucial\cite{Luo2016Nov}. Similar conclusions were drawn from recent Self-Interaction-Error-corrected DFT calculations\cite{Zhang2019Jan}. However, the complete nucleation picture drawn at the atomic scale remains unclear in the case of titania since very few finite temperature MD simulations were performed\cite{Yang2019,Mavracic2018Jun,Alderman2014Sep} and no a single one focused on polymorph selection because short ab initio MD or classical empirical force fields were employed.

\section{Conclusions}

Despite the key technological importance of ZnO in many applications, its crystallization features are at present relatively poorly understood. This is due to overwhelming computational effort necessary for dedicated simulations, which excludes a direct use of DFT-based methods and requires fine interatomic potentials, able to correctly account for the structural diversity of ZnO as well as for its behaviour at finite and high temperatures.

With the goal to observe crystal nucleation in a homogeneous ZnO liquid phase and to follow the subsequent crystal growth, we have constructed and validated a new robust machine learning interatomic potential suitable for large-scale simulations (nanosecond molecular dynamics of several thousands of atoms). The potential is based on physically-motivated mathematical formulation and a constrained LassoLars method is used to identify and adjust its pertinent parameters. The training database was composed of DFT-GGA results on a variety of ordered crystalline structures while also including surface and disordered configurations. We show that the new interaction potential successfully reproduces the delicate structural and energetic DFT-GGA characteristics of six of the most stable ZnO bulk polymorphs as well as of their low-index surfaces. The potential was also successfully tested on completely untrained structures made of nanoparticles in different crystal phases and was shown to reproduce the vibrational characteristics and free energy behaviour of the most stable bulk polymorphs. Importantly, while not trained in this regime, it also correctly accounts for the structural characteristics of high temperature liquid ZnO such that they become quasi-indistinguishable from the DFT ones. These extensive testings demonstrate the reliability and the transferability of the PLIP model which was then employed to study the crystallization of zinc oxide. 

In particular, we have performed molecular dynamics simulations on freezing of bulk ZnO liquid as well as equilibrium calculations of undercooled ZnO liquid crystallization under a constant temperature and pressure conditions. Analysis of the crystalline phase in such complex structural landscape was made possible by employing a data-driven approach based on Steinhardt’s parameters which enabled us to access the nucleation of the initial zinc oxide crystal at the atomistic level and to follow its behavior upon further growth. We have systematically observed two different scenarios which are both in contradiction to classical nucleation theory: (1) Prenucleation clusters made of two coexisting phases, WRZ in the core and BCT in the shell and (2) Two-step nucleation process with metastable BCT emerging first and turning into stable WRZ afterwards. Similar nucleation scenarios were previously observed in much simpler systems including Lennard-Jones, hard-sphere, and pure metals. In our case, the combination of the Physical LassoLars interaction potential that allows for a quantum-accurate modeling of zinc oxide with our data-driven approach for a refined structural analysis of crystal phases leads us to confirm previous predictions of non-classical nucleation processes.\cite{Jehannin2019Jul,Lee2016Jun,DeYoreo2020} Moreover, dedicated calculations of ZnO free energy within a newly developed automated nonequilibrium thermodynamic integration method showed the existence of a stability reversal between wurtzite and BCT polymorphs at high temperatures which advocates for the role of thermodynamic in explaining the observed non-classical nucleation picture.

As a perspective, we note that the proposed work paves the way for more complex numerical studies that would account the influence of pressure, finite liquid size and/or presence of surfaces, substrate, and/or temperature gradient, relevant for fabrication of technologically important ZnO nanostructures and should help understanding the relationship between their characteristics and the conditions of their early growth stages. 

\section*{Acknowledgement}

JL acknowledges financial support of the Fonds de la Recherche Scientifique - FNRS. Computational resources have been provided by the Consortium des Equipements de Calcul Intensif (CECI), by the F\'ed\'eration Lyonnaise de Mod\'elisation et Sciences Numériques (FLMSN) and by the Regional Computer Center CALMIP in Toulouse. JL thanks David Mora Fonz, Sarath Menon and Anup Pandey for fruitful discussions.

\section*{Data availability}

The data generated in this study are available on Github [link will be given upon acceptance]

%
%\newpage
%\clearpage
%
%\includepdf[landscape=false]{SI.pdf}
\end{document}